\documentclass[aps,prd,twocolumn,nofootinbib,,superscriptaddress,showkeys]{revtex4}
\usepackage{amssymb,amsmath,graphicx,enumerate}
\usepackage{subfigure}
\usepackage{dcolumn,booktabs}
\usepackage[dvipdfm,colorlinks=true,citecolor=blue,pdfstartview=FitH]{hyperref}
\usepackage{color,framed} 
\definecolor{shadecolor}{rgb}{0,0,1} 

\def\bea{\begin{equation}}
\def\eea{\end{equation}}
\newcommand{\rt}{Regge trajectory}
\newcommand{\rts}{Regge trajectories}

\newcommand{\trs}{trajectories}
\newcommand{\bfr}{{\bf r}}
\newcommand{\bfp}{{\bf p}}
\newcommand{\bfpa}{{|\bf p|}}
\newcommand{\gev}{{\rm GeV}}
\newcommand{\mev}{{\rm MeV}}
\newcommand{\cltb}{$\bar{3}_c$}
\newcommand{\cltba}{\bar{3}_c}

\newcommand{\dqs}{$(qq')$}
\newcommand{\qp}{{q^{\prime}}}

\newcommand{\qpp}{{q^{\prime\prime}}}

\newcommand{\qqs}{{[qq^{\prime}]}}
\newcommand{\qqb}{\{qq^{\prime}\}}
\newcommand{\dhb}{doubly heavy baryon}
\newcommand{\dhbs}{doubly heavy baryons}

\begin{document}
\title{$\lambda$ and $\rho$ trajectories for the doubly heavy baryons in the diquark picture}
\author{He Song}
\email{songhe\_22@163.com}
\affiliation{School of Physics and Information Engineering, Shanxi Normal University, Taiyuan 030031, China}
\author{Jia-Qi Xie}
\email{1462718751@qq.com}
\affiliation{School of Physics and Information Engineering, Shanxi Normal University, Taiyuan 030031, China}
\author{Xin-Ru Liu}
\email{1170394732@qq.com}
\affiliation{School of Physics and Information Engineering, Shanxi Normal University, Taiyuan 030031, China}
\author{Jiao-Kai Chen}
\email{chenjk@sxnu.edu.cn, chenjkphy@outlook.com (corresponding author)}
\affiliation{School of Physics and Information Engineering, Shanxi Normal University, Taiyuan 030031, China}
\begin{abstract}
We present the explicit form of the Regge trajectory relations for the doubly heavy baryons $\Xi_{QQ'}$ and $\Omega_{QQ'}$ $(Q,Q'=b,c)$ in the diquark picture.
Using the derived Regge trajectory relations, we estimate the masses of the $\lambda$-excited states and the $\rho$-excited states, which are consistent with other theoretical predictions.
Both the $\lambda$-trajectories and $\rho$-trajectories are discussed. We show that the $\rho$-trajectories behave differently from the $\lambda$-trajectories. Specifically, the $\rho$-trajectories behave as $M{\sim}x_{\rho}^{2/3}$ $(x_{\rho}=n_r,l)$, whereas the $\lambda$-trajectories follow $M{\sim}x_{\lambda}^{1/2}$ $(x_{\lambda}=N_r,L)$.
By using the obtained relations, the baryon Regge trajectory provides a straightforward and easy method for estimating the spectra of both the $\lambda$-excited states and $\rho$-excited states.
\end{abstract}

\maketitle


\section{Introduction}

In the diquark picture, the {\dhbs} $\Xi_{QQ'}$ and $\Omega_{QQ'}$ $(Q,Q'=b,\,c)$ are composed of a doubly heavy diquark $(QQ')$ and a light quark $q$ $(q=u,\,d,\,s)$. The {\dhbs} \cite{Klempt:2009pi,Chen:2022asf,
Richard:1992uk} have been studied by a wide variety of approaches. These approaches include the quark model \cite{Karliner:2018hos}, the nonrelativistic potential model \cite{Gershtein:2000nx,Chen:2022fye}, the relativistic quark model \cite{Ebert:2002ig}, the hypercentral constituent quark model \cite{Shah:2017liu,Shah:2016vmd}, the relativized quark model \cite{Lu:2017meb}, Lattice QCD \cite{Mathur:2018epb,Mohanta:2019mxo}, the Bethe-Salpeter equation \cite{Li:2019ekr,Yu:2018com}, QCD sum rules \cite{Tang:2011fv,Wang:2018lhz,Aliev:2019lvd}, the nonrelativistic quark model \cite{Albertus:2006wb}, and the Faddeev equation \cite{Gutierrez-Guerrero:2024him}, among others.

The {\rt} is one of the effective approaches for studying hadron spectra \cite{Regge:1959mz,Chew:1962eu,Nambu:1978bd,Inopin:1999nf,Brisudova:1999ut,
Chen:2014nyo,Wei:2016jyk,Nielsen:2018uyn,Sonnenschein:2018fph,Song:2022csw,
MartinContreras:2020cyg,MartinContreras:2023oqs,Roper:2024ovj,Sergeenko:1994ck,
Afonin:2014nya,Burns:2010qq,Ishida:1994pf,Costa-Silva:2024dlq,Guo:2024nrf,
A:2023bxv,Chen:2018hnx,Chen:2018bbr,Chen:2018nnr,Song:2024bkj}. In previous works \cite{Ebert:2002ig,Wei:2016jyk,Shah:2017liu,Shah:2016vmd,Ishida:1994pf}, the {\rt} approach has been applied to calculate the masses of the $\lambda$-mode states or to fit data obtained through other methods.
In Ref. \cite{Xie:2024lfo}, the explicit form of the {\rt} relations for the triply heavy bottom-charm baryons is presented and the masses of the $\lambda$-mode excited states and the $\rho$-mode excited states are estimated by the given {\rt} relations. It is noted that the form of the {\rt} relations and the parameters used are universal for diquarks, mesons, baryons and tetraquarks \cite{Xie:2024lfo,Feng:2023txx,Chen:2023djq,Chen:2023web}.
Following Ref. \cite{Xie:2024lfo}, this work provides the explicit form of the {\rt} relations for the {\dhbs} in the diquark picture by employing the diquark {\rt} relation \cite{Feng:2023txx}. Both the $\lambda$-trajectories and the $\rho$-trajectories are presented, and the corresponding masses are estimated using the obtained {\rt} relations.

The paper is organized as follows: In Sec. \ref{sec:rgr}, the {\rt} relations for {\dhbs} are obtained from the spinless Salpeter equation. In Sec. \ref{sec:rtdiquark}, we investigate the {\rts} for the {\dhbs} and the masses of the excited states are provided. The conclusions are summarized in Sec. \ref{sec:conc}.

\section{{\rt} relations}\label{sec:rgr}
In this section, the {\rt} relations for the {\dhbs} are obtained from the spinless Salpeter equation.

\subsection{Preliminary}\label{subsec:prelim}

\begin{figure}[!phtb]
\centering
\includegraphics[width=0.25\textheight]{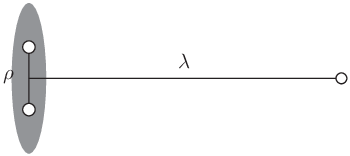}
\caption{Schematic diagram of the baryons in the diquark-quark picture.}\label{fig:tr}
\end{figure}

In the diquark picture, baryons consist of a quark in color $3_c$ and a diquark in color $\bar{3}_c$, as shown in Fig. \ref{fig:tr}. $\rho$ separates the quarks in the diquark, and $\lambda$ separates the quark and the diquark. There are two excited modes: the $\rho$-mode involves the radial and orbital excitations in the diquark, and the $\lambda-$mode involves the radial or orbital excitations between the quark and diquark. As a result, there are two series of {\rts}: one series of $\rho$-{\trs} and one series of $\lambda$-{\trs}. However, various mixings \cite{Gershtein:2000nx,Ebert:2002ig,Faustov:2021qqf} often occur, leading to complexities in the $\rho$-{\trs} and the $\lambda$-{\trs}.

\begin{table}[!phtb]
\caption{The completely antisymmetric states for the diquarks in {\cltb} \cite{Feng:2023txx}. $j_d$ is the spin of the diquark {\dqs}, $s_d$ denotes the total spin of two quarks, $l$ represents the orbital angular momentum. $n=n_r+1$, $n_r$ is the radial quantum number, $n_r=0,1,2,\cdots$. }  \label{tab:dqstates}
\centering
\begin{tabular*}{0.49\textwidth}{@{\extracolsep{\fill}}ccccc@{}}
\hline\hline
 Spin of diquark & Parity  &  Wave state  &  Configuration    \\
( $j_d$ )          & $(j_d^P)$ & $(n^{2s_d+1}l_{j_d})$  \\
\hline
$j_d=0$              & $0^+$   & $n^1s_0$         & $[qq']^{{\cltba}}_{n^1s_0}$ \\
                 & $0^-$   & $n^3p_0$         & $[qq']^{{\cltba}}_{n^3p_0}$       \\
$j_d=1$              & $1^+$   & $n^3s_1$, $n^3d_1$   & $\{qq'\}^{{\cltba}}_{n^3s_1}$,\;    $\{qq'\}^{{\cltba}}_{n^3d_1}$\\
                 & $1^-$   & $n^1p_1$, $n^3p_1$   &
$\{qq'\}^{{\cltba}}_{n^1p_1}$,\; $[qq']^{{\cltba}}_{n^3p_1}$ \\
$j_d=2$              & $2^+$   & $n^1d_2$, $n^3d_2$         &  $[qq']^{{\cltba}}_{n^1d_2}$,\; $\{qq'\}^{{\cltba}}_{n^3d_2}$\\
                 & $2^-$   & $n^3p_2$, $n^3f_2$       &
 $[qq']^{{\cltba}}_{n^3p_2}$,\; $[qq']^{{\cltba}}_{n^3f_2}$       \\
$\cdots$         & $\cdots$ & $\cdots$               & $\cdots$  \\
\hline\hline
\end{tabular*}
\end{table}

In the diquark picture, the state of a baryon is denoted as
\bea\label{tetnot}
\left((qq')_{n^{2s_d+1}l_{j_d}}
{\qpp}\right)_{N^{2j+1}L_J},
\eea
where the superscripts {\cltb}, $3_c$ and $1_c$ are omitted. The diquark $(qq')$ is either $\{qq'\}$ or $[qq']$, where $\{qq'\}$ and $[qq']$ indicate the permutation symmetric and antisymmetric flavor wave functions, respectively. The completely antisymmetric states for the diquarks in {\cltb} are listed in Table \ref{tab:dqstates}. $N=N_{r}+1$, where $N_{r}=0,\,1,\,\cdots$. $n=n_{r}+1$, where $n_{r}=0,\,1,\,\cdots$. $N_r$ and $n_{r}$ are the radial quantum numbers of the baryon and diquark, respectively.
$\vec{J}=\vec{L}+\vec{j}$, $\vec{j}=\vec{j}_d+\vec{s}_{\qpp}$, $\vec{j}_d=\vec{s}_d+\vec{l}$.
$\vec{J}$, $\vec{j}_d$ and $\vec{s}_{\qpp}$ are the spins of baryon, diquark and quark $q^{\prime\prime}$, respectively. $\vec{j}$ is the summed spin of diquark and quark in the baryon. $L$ and $l$ are the orbital quantum numbers of baryon and diquark, respectively. $\vec{s}_{d}$ is the summed spin of quarks in the diquark.

\subsection{Spinless Salpeter equation}
The spinless Salpeter equation \cite{Godfrey:1985xj,Ferretti:2019zyh,Bedolla:2019zwg,Durand:1981my,Durand:1983bg,Lichtenberg:1982jp,Jacobs:1986gv} reads as
\begin{eqnarray}\label{qsse}
M\Psi_{d,b}({\bfr})=\left(\omega_1+\omega_2\right)\Psi_{d,b}({\bfr})+V_{d,b}\Psi_{d,b}({\bfr}),
\end{eqnarray}
where $M$ is the bound state mass (diquark or baryon). $\Psi_{d,b}({\bfr})$ are the diquark wave function and the baryon wave function, respectively. $V_{d,b}$ denotes the diquark potential and the baryon potential, respectively (see Eq. (\ref{potv})). $\omega_1$ is the relativistic energy of constituent $1$ (quark or diquark), and $\omega_2$ is of constituent $2$ (quark),
\bea\label{omega}
\omega_i=\sqrt{m_i^2+{\bf p}^2}=\sqrt{m_i^2-\Delta}\;\; (i=1,2).
\eea
Here, $m_1$ and $m_2$ are the effective masses of constituent $1$ and $2$, respectively.

When using diquark in multiquark systems, it is necessary to consider the interactions between quark and quark, diquark and quark, and diquark and diquark. In Ref. \cite{Faustov:2021hjs}, these interactions are constructed using the off-mass-shell scattering amplitude, which is projected onto positive energy states.
Alternatively, the interactions can be established by extending the interactions of the quark-antiquark system to the quark-quark system, and then to the diquark-antidiquark systems or the diquark-quark systems \cite{Lundhammar:2020xvw}.
Moreover, the effect of the finite size of diquark is treated differently in various studies. In Ref. \cite{Faustov:2021hjs}, the size of diquark is taken into account through corresponding form factors. In some cases, however, diquark is taken as being pointlike \cite{Ferretti:2019zyh,Lundhammar:2020xvw}, which is the approximation adopted in the present work.

Following Refs. \cite{Ferretti:2019zyh,Bedolla:2019zwg,Ferretti:2011zz,Eichten:1974af,
Chen:2023web,Chen:2023djq}, we employ the potential
\begin{align}\label{potv}
V_{d,b}&=-\frac{3}{4}\left[V_c+{\sigma}r+C\right]
\left({\bf{F}_i}\cdot{\bf{F}_j}\right)_{d,b},
\end{align}
where $V_c\propto{1/r}$ is a color Coulomb potential or a Coulomb-like potential arising from one-gluon-exchange. $\sigma$ is the string tension, and $C$ is a fundamental parameter \cite{Gromes:1981cb,Lucha:1991vn}. The part in the bracket is the Cornell potential \cite{Eichten:1974af}. ${\bf{F}_i}\cdot{\bf{F}_j}$ is the color-Casimir, with
\bea\label{mrcc}
\langle{(\bf{F}_i}\cdot{\bf{F}_j})_{d}\rangle=-\frac{2}{3},\quad
\langle{(\bf{F}_i}\cdot{\bf{F}_j})_{b}\rangle=-\frac{4}{3}.
\eea

\subsection{{\rt} relations for the heavy-light and heavy-heavy systems}
In this subsection, we present the {\rt} relations for the heavy-heavy systems and the heavy-light systems.

For the heavy-heavy systems, $m_{1},m_2{\gg}{\bfpa}$, Eq. (\ref{qsse}) reduces to
\begin{eqnarray}\label{qssenrr}
M\Psi_{d,t}({\bfr})&=&\left[(m_1+m_2)+\frac{{\bfp}^2}{2\mu}\right]\Psi_{d,t}({\bfr})\nonumber\\
&&+V_{d,t}\Psi_{d,t}({\bfr}),
\end{eqnarray}
where
\bea\label{rdmu}
\mu=m_1m_2/(m_1+m_2).
\eea
By employing the Bohr-Sommerfeld quantization approach \cite{Brau:2000st} and using Eqs. (\ref{potv}) and (\ref{qssenrr}), we obtain the parameterized relation \cite{Chen:2022flh,Chen:2021kfw}
\bea\label{massform}
M=m_R+\beta_x(x+c_{0x})^{2/3}\;(x=l,\,n_r,\,L,\,N_r)
\eea
with
\bea\label{rtft}
m_R=m_1+m_2+C',
\eea
where
\bea\label{cprime}
C'=\left\{\begin{array}{cc}
C/2, & \text{diquarks}, \\
C, & \text{baryons}.
\end{array}\right.
\eea
\bea\label{sigma}
\sigma'=\left\{\begin{array}{cc}
\sigma/2, & \text{diquarks}, \\
\sigma, & \text{baryons}.
\end{array}\right.
\eea
$\beta_x$ reads
\bea\label{parabm}
\beta_x=c_{fx}c_xc_c.
\eea
$c_c$ and $c_x$ are
\bea
c_c=\left(\frac{\sigma'^2}{\mu}\right)^{1/3},\; c_l=\frac{3}{2},\; c_{n_r}=\frac{\left(3\pi\right)^{2/3}}{2}.
\eea
Both $c_{fl}$ and $c_{fn_r}$ are equal theoretically to one but are determined through fitting in practice.
In Eq. (\ref{massform}), $m_1$, $m_2$, $c_x$ and $\sigma$ are universal. $c_{0x}$ are determined by fitting.

For the heavy-light systems ($m_1\to\infty$ and $m_2\to0$), Eq. (\ref{qsse}) simplifies to
\begin{eqnarray}\label{qssenr}
M\Psi_{d,t}({\bfr})=\left[m_1+{\bfpa}+V_{d,t}\right]\Psi_{d,t}({\bfr}).
\end{eqnarray}
By employing the Bohr-Sommerfeld quantization approach \cite{Brau:2000st}, the parameterized formula can be obtained and is expressed as \cite{Chen:2022flh,Chen:2021kfw}
\bea\label{rtmeson}
M=m_R+\beta_x\sqrt{x+c_{0x}}\;(x=l,\,n_r,\,L,\,N_r).
\eea
The parameters in Eq. (\ref{rtmeson}) read as
\bea\label{massformhl}
c_{c}=\sqrt{\sigma'},\; c_l=2,\; c_{n_r}=\sqrt{2\pi}.
\eea
There are different ways to include the mass of the light constituent \cite{Selem:2006nd,Nielsen:2018uyn,Sonnenschein:2018fph,
MartinContreras:2020cyg,Chen:2023cws,Chen:2023ngj,Chen:2022flh,Chen:2014nyo,
Afonin:2014nya,Sergeenko:1994ck}. We use Eqs. (\ref{rtmeson}) and (\ref{rtft}) \cite{Xie:2024lfo,Chen:2023cws,Chen:2023ngj,Xie:2024dfe}.

\begin{table}[!phtb]
\caption{The coefficients for the heavy-heavy systems (HHS) and the heavy-light systems (HLS).}  \label{tab:eparam}
\centering
\begin{tabular*}{0.47\textwidth}{@{\extracolsep{\fill}}ccc@{}}
\hline\hline
                   & HHS &  HLS   \\
\hline
$\nu$    & $2/3$ & $1/2$    \\
$c_c$    & $\left({\sigma'^2}/{\mu}\right)^{1/3}$    & $\sqrt{\sigma'}$  \\
$c_{l,\,L}$    & $3/2$ & $2$   \\
$c_{n_r,\,N_r}$ & ${\left(3\pi\right)^{2/3}}/{2}$      & $\sqrt{2\pi}$   \\
\hline
\hline
\end{tabular*}
\end{table}

When Eqs. (\ref{massform}) and (\ref{rtft}) are applied to discuss the heavy-heavy systems, and Eqs. (\ref{rtmeson}) and (\ref{rtft}) are employed to discuss the heavy-light systems, we can summarise Eqs. (\ref{massform}), (\ref{rtft}) and (\ref{rtmeson}) into a general form of the {\rts} \cite{Xie:2024lfo,Chen:2022flh,Xie:2024dfe}
\begin{align}\label{massfinal}
M=&m_R+\beta_x(x+c_{0x})^{\nu}\,\,(x=l,\,n_r,\,L,\,N_r),\nonumber\\
m_R=&m_1+m_2+C',\quad \beta_x=c_{fx}c_xc_{c},
\end{align}
where ${\nu}$, $c_x$ and $c_{c}$ are listed in Table \ref{tab:eparam}. $c_{fx}$ are theoretically equal to one but are determined through fitting in practice. $c_{0x}$ vary with different {\rts}.

Eq. (\ref{massfinal}) can be employed to discuss various systems including the heavy-heavy systems, the heavy-light systems, and the light-light systems: diquarks, mesons, baryons, triquarks, and tetraquarks \cite{Chen:2023djq,Chen:2023ngj,Chen:2023web,Song:2024bkj}.
The parameter values used are universal for both the heavy-heavy systems and the heavy-light systems \cite{Chen:2023cws,Feng:2023txx}, and are applied universally to various types of bound states \cite{Song:2024bkj,Xie:2024lfo,
Chen:2023cws,Feng:2023txx,Xie:2024dfe}.
However, for the light systems, the parameter values require adjustment to achieve consistent results \cite{Chen:2023ngj}.
The general form in Eq. (\ref{massfinal}) is provisional and remains open to refinement because there are different methods for including the masses of the light constituents. Further theoretical and experimental studies are required to improve its reliability.

\subsection{{\rt} relations for the doubly heavy baryons $(QQ')q$ }
For a baryon $(QQ')q$ composed of one doubly heavy diquark and one light quark, according to Eq. (\ref{massfinal}), we have the {\rt} relations
\begin{align}\label{t2q}
M&=m_{R{\lambda}}+\beta_{x_{\lambda}}(x_{\lambda}+c_{0x_{\lambda}})^{1/2}\;(x_{\lambda}=L,\,N_r),\nonumber\\
M_{\rho}&=m_{R\rho}+\beta_{x_{\rho}}(x_{\rho}+c_{0x_{\rho}})^{2/3}\;(x_{\rho}=l,\,n_{r}),
\end{align}
where
\begin{align}\label{pa2qQ}
m_{R{\lambda}}&=M_{\rho}+m_{q}+C,\nonumber\\
m_{R\rho}&=m_{Q}+m_{Q'}+C/2,\nonumber\\
\beta_{L}&=\sqrt{4\sigma}c_{fL},\; \beta_{N_r}=\sqrt{2\pi\sigma}c_{fN_r},\;
\mu_{\rho}=\frac{m_{Q}m_{Q^{\prime}}}{m_{Q}+m_{Q^{\prime}}},\nonumber\\
\beta_{l}&=\frac{3}{2}\left(\frac{\sigma^2}{4\mu_{\rho}}\right)^{1/3}c_{fl},\; \beta_{n_r}=\frac{(3\pi)^{2/3}}{2}\left(\frac{\sigma^2}{4\mu_{\rho}}\right)^{1/3}c_{fn_r}.
\end{align}
In Eq. (\ref{t2q}), $M$ is the mass of baryon, and $M_{\rho}$ is the mass of diquark. The second relation in Eq. (\ref{t2q}) is the diquark {\rt} relation and is used to calculate the diquark masses. The first relation in Eq. (\ref{t2q}), along with the other relations in Eqs. (\ref{t2q}) and (\ref{pa2qQ}), is employed to calculate baryon masses.

According to Eqs. (\ref{t2q}) and (\ref{pa2qQ}), we have for a {\dhb} $(QQ')q$
\bea
M=M_{\rho}+m_{q}+C+\beta_{x_{\lambda}}(x_{\lambda}+c_{0x_{\lambda}})^{1/2}
\eea
when the internal structure of the diquark is not considered, treating $M_{\rho}$ as an input parameter. When the diquark is considered as a bound state composed of two heavy quarks \cite{Feng:2023txx}, we have
\begin{align}\label{combrt}
M=&m_{Q}+m_{Q'}+m_{q}+\frac{3C}{2}\nonumber\\
&+\beta_{x_{\lambda}}(x_{\lambda}+c_{0x_{\lambda}})^{1/2}
+\beta_{x_{\rho}}(x_{\rho}+c_{0x_{\rho}})^{2/3}
\end{align}
from Eqs. (\ref{t2q}) and (\ref{pa2qQ}).
From Eq. (\ref{combrt}), it is evident that there are two series of {\rts} for {\dhbs}: the $\lambda$-trajectories and the $\rho$-trajectories. These two series of {\rts} exhibit distinct behaviors, $M{\sim}x_{\lambda}^{1/2}$ and $M{\sim}x_{\rho}^{2/3}$.

\section{{\rts} for the doubly heavy baryons}\label{sec:rtdiquark}

In this section, we estimate the masses of {\dhbs} $\Xi_{cc}$, $\Xi_{bc}$, $\Xi_{bb}$, $\Omega_{cc}$, $\Omega_{bc}$ and $\Omega_{bb}$. The $\lambda$-trajectories and $\rho$-trajectories are investigated by using Eqs. (\ref{t2q}) and (\ref{pa2qQ}) or Eqs. (\ref{combrt}) and (\ref{pa2qQ}).

\subsection{Parameters}

The parameter values are listed in Table \ref{tab:parmv}. The values of $m_u$, $m_d$, $m_s$, $m_b$, $m_c$, $\sigma$ and $C$ are taken directly from \cite{Faustov:2021qqf}, and are universal for both the $\lambda$-mode and $\rho$-mode. The parameters $c_{fx}$ and $c_{0x}$ for the $\rho$-mode are obtained by fitting the {\rts} for doubly heavy mesons. $c_{fx}$ are universal for all doubly heavy diquark {\rts}, whereas $c_{0x}$ varies with different diquark {\rts} \cite{Feng:2023txx}.
For the $\lambda$-mode of the {\dhbs} $(QQ'){q}$, the parameters $c_{fx}$ and $c_{0x}$  are determined using the relations in Eq. (\ref{fitcfxl}).
These relations are used as a provisional method until a more refined approach is developed. The validity of this method can be assessed by comparing the fitted results for doubly heavy baryons with the theoretical values obtained through other approaches.

\begin{table}[!phtb]
\caption{The values of parameters \cite{Faustov:2021qqf,Feng:2023txx}.}  \label{tab:parmv}
\centering
\begin{tabular*}{0.45\textwidth}{@{\extracolsep{\fill}}cc@{}}
\hline\hline
          & $m_{u}=m_{d}=0.33\; {\gev}$, \; $m_s=0.5\; {\gev}$, \\
          & $m_{c}=1.55\; {\gev}$, \; $m_b=4.88\; {\gev}$, \\
          & $\sigma=0.18\; {\gev^2}$,\; $C=-0.3\; {\gev}$, \\
          & $c_{fn_{r}}=1.0$,\; $c_{fl}=1.17$        \\
$(cc)$    & $c_{0n_{r}}(1^3s_1)=0.205$,\quad $c_{0{l}}(1^3s_1)=0.337$,\\
$(bc)$    & $c_{0n_{r}}(1^3s_1)=0.182$,\quad $c_{0{l}}(1^3s_1)=0.257$,\\
          & $c_{0n_{r}}(1^1s_0)=0.107$,\quad $c_{0{l}}(1^1s_0)=0.169$,\\
$(bb)$    & $c_{0n_{r}}(1^3s_1)=0.01$,\quad  $c_{0{l}}(1^3s_1)=0.001$.\\
\hline
\hline
\end{tabular*}
\end{table}

\begin{table}[!phtb]
\caption{The spin-averaged masses of the $\lambda$-excited states of $(bb)q$ and $(cc)q$ (in ${\gev}$). $q=u,\,s$. The notation in Eq. (\ref{tetnot}) is rewritten as $|n^{2s_d+1}l_{j_d},N^{2j+1}L_J\rangle$. And $|n^{2s_d+1}l_{j_d},NL\rangle$ denotes the spin-averaged states. Eqs. (\ref{t2q}), (\ref{pa2qQ}) and (\ref{fitcfxl}) are used.}  \label{tab:masslambda}
\centering
\begin{tabular*}{0.48\textwidth}{@{\extracolsep{\fill}}ccccc@{}}
\hline\hline
  $|n^{2s_d+1}l_{j_d},NL\rangle$ & $(cc)u$ & $(cc)s$ & $(bb)u$ & $(bb)s$ \\
\hline
 $|1^3s_1, 1S\rangle$  &3.67    &3.83   &10.22    &10.37   \\
 $|1^3s_1, 2S\rangle$  &4.32    &4.46   &10.85    &10.98     \\
 $|1^3s_1, 3S\rangle$  &4.73    &4.86   &11.24    &11.36      \\
 $|1^3s_1, 4S\rangle$  &5.04    &5.17   &11.56    &11.67      \\
 $|1^3s_1, 5S\rangle$  &5.32    &5.43   &11.83    &11.93      \\
\hline
 $|1^3s_1, 1S\rangle$  &3.69    &3.84   &10.22    &10.37      \\
 $|1^3s_1, 1P\rangle$  &4.14    &4.28   &10.66    &10.79    \\
 $|1^3s_1, 1D\rangle$  &4.44    &4.58   &10.95    &11.08     \\
 $|1^3s_1, 1F\rangle$  &4.69    &4.81   &11.20    &11.32      \\
 $|1^3s_1, 1G\rangle$  &4.90    &5.02   &11.40    &11.52      \\
 $|1^3s_1, 1H\rangle$  &5.08    &5.20   &11.59    &11.70    \\
\hline\hline
\end{tabular*}
\end{table}

\begin{table}[!phtb]
\caption{Same as Table \ref{tab:masslambda} except for baryons $(bc)q$. }  \label{tab:masslambdabc}
\centering
\begin{tabular*}{0.48\textwidth}{@{\extracolsep{\fill}}ccc@{}}
\hline\hline
  $|n^{2s_d+1}l_{j_d},NL\rangle$ & $(bc)u$ & $(bc)s$ \\
\hline
 $|1^3s_1, 1S\rangle$  &7.00    &7.15      \\
 $|1^3s_1, 2S\rangle$  &7.63    &7.77      \\
 $|1^3s_1, 3S\rangle$  &8.03    &8.15      \\
 $|1^3s_1, 4S\rangle$  &8.35    &8.46      \\
 $|1^3s_1, 5S\rangle$  &8.62    &8.72      \\
\hline
 $|1^1s_0, 1S\rangle$  &6.96    &7.11      \\
 $|1^1s_0, 2S\rangle$  &7.59    &7.73      \\
 $|1^1s_0, 3S\rangle$  &7.99    &8.11      \\
 $|1^1s_0, 4S\rangle$  &8.31    &8.42      \\
 $|1^1s_0, 5S\rangle$  &8.58    &8.68      \\
\hline
 $|1^3s_1, 1S\rangle$  &7.01    &7.16      \\
 $|1^3s_1, 1P\rangle$  &7.45    &7.58      \\
 $|1^3s_1, 1D\rangle$  &7.74    &7.87      \\
 $|1^3s_1, 1F\rangle$  &7.99    &8.11      \\
 $|1^3s_1, 1G\rangle$  &8.19    &8.31      \\
 $|1^3s_1, 1H\rangle$  &8.38    &8.49      \\
 \hline
 $|1^1s_0, 1S\rangle$  &6.96    &7.12     \\
 $|1^1s_0, 1P\rangle$  &7.40    &7.54      \\
 $|1^1s_0, 1D\rangle$  &7.70    &7.83      \\
 $|1^1s_0, 1F\rangle$  &7.94    &8.07      \\
 $|1^1s_0, 1G\rangle$  &8.15    &8.27    \\
 $|1^1s_0, 1H\rangle$  &8.34    &8.45     \\
\hline\hline
\end{tabular*}
\end{table}

\begin{figure*}[!phtb]
\centering
\subfigure[]{\label{subfigure:cfa}\includegraphics[scale=0.48]{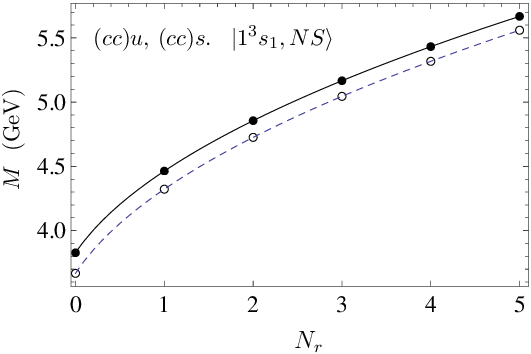}}
\subfigure[]{\label{subfigure:cfa}\includegraphics[scale=0.48]{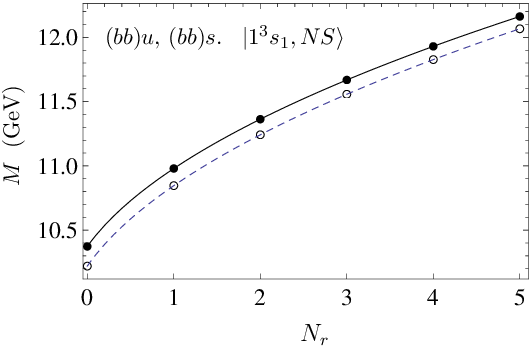}}
\subfigure[]{\label{subfigure:cfa}\includegraphics[scale=0.48]{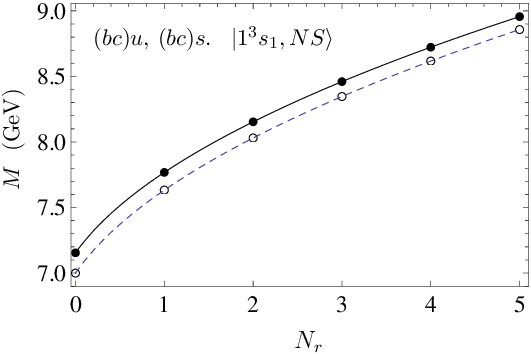}}
\subfigure[]{\label{subfigure:cfa}\includegraphics[scale=0.48]{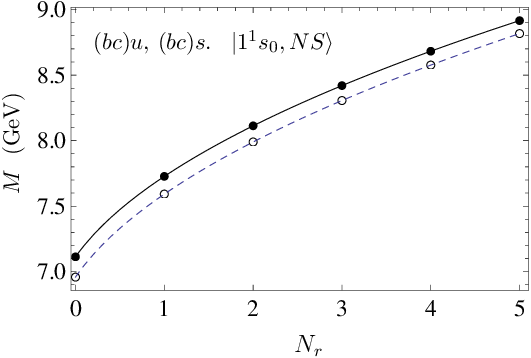}}
\subfigure[]{\label{subfigure:cfa}\includegraphics[scale=0.48]{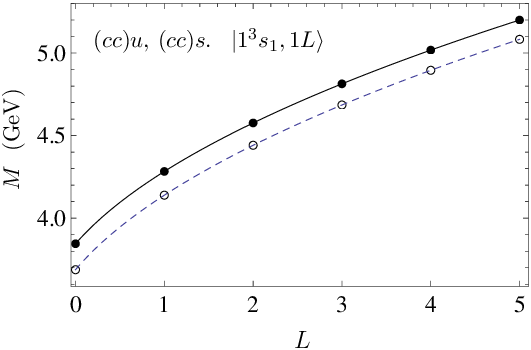}}
\subfigure[]{\label{subfigure:cfa}\includegraphics[scale=0.48]{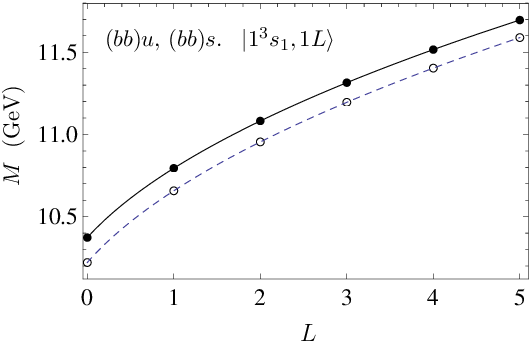}}
\subfigure[]{\label{subfigure:cfa}\includegraphics[scale=0.48]{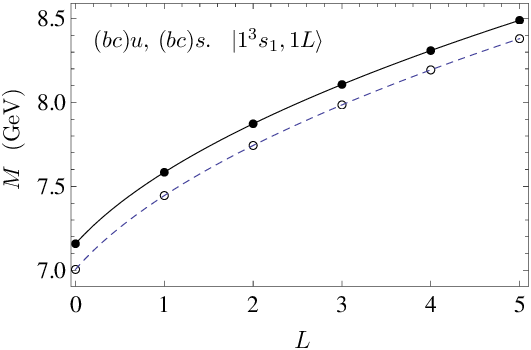}}
\subfigure[]{\label{subfigure:cfa}\includegraphics[scale=0.48]{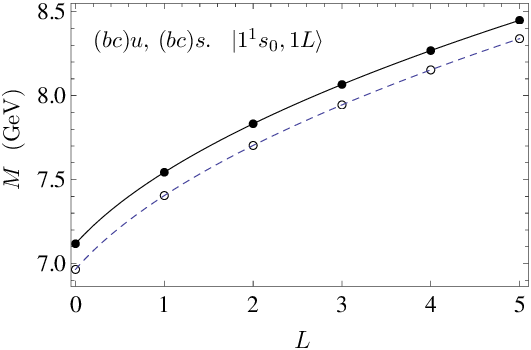}}
\caption{The radial and orbital $\lambda$-trajectories for the {\dhbs} $(QQ')q$. Circles and the dashed lines are for the baryons $(QQ')u$ containing an up quark. Dotes and the black lines are for the baryons $(QQ')s$ containing a strange quark. $M$ is the baryon mass. $N=N_r+1$, $N_r$ is the radial quantum number for the $\lambda$-mode. $L$ is the orbital quantum number for the $\lambda$-mode. Data are listed in Tables \ref{tab:masslambda} and \ref{tab:masslambdabc}.}\label{fig:brho}
\end{figure*}

\begin{table}[!htbp]
\caption{Same as Table \ref{tab:masslambda} except for the $\rho$-excited states. $\times$ denotes the nonexist states.} \label{tab:massrho}
\centering
\begin{tabular*}{0.48\textwidth}{@{\extracolsep{\fill}}ccccc@{}}
\hline\hline
  $|n^{2s_d+1}l_{j_d},NL\rangle$ & $(cc)u$ & $(cc)s$ & $(bb)u$ & $(bb)s$ \\
\hline
 $|1^3s_1, 1S\rangle$  &3.67    &3.83    &10.22    &10.37 \\
 $|2^3s_1, 1S\rangle$  &4.06    &4.22    &10.54    &10.69 \\
 $|3^3s_1, 1S\rangle$  &4.34    &4.49    &10.74    &10.89 \\
 $|4^3s_1, 1S\rangle$  &4.57    &4.73    &10.90    &11.05 \\
 $|5^3s_1, 1S\rangle$  &4.79    &4.94    &11.05    &11.20 \\
\hline
 $|1^3s_1, 1S\rangle$            &3.68    &3.84    &10.21  &10.36 \\
 $|1^3p_2, 1S\rangle$$(\times)$  &3.97    &4.13  &10.47    &10.62 \\
 $|1^3d_3, 1S\rangle$            &4.18    &4.34  &10.62    &10.77 \\
 $|1^3f_4, 1S\rangle$$(\times)$  &4.37    &4.52  &10.75    &10.90 \\
 $|1^3g_5, 1S\rangle$            &4.53    &4.69  &10.87    &11.02  \\
 $|1^3h_6, 1S\rangle$$(\times)$  &4.69    &4.84  &10.97    &11.12 \\
\hline\hline
\end{tabular*}
\end{table}

\begin{table}[!htbp]
\caption{Same as Table \ref{tab:masslambdabc} except for the $\rho$-excited states. } \label{tab:massrhobc}
\centering
\begin{tabular*}{0.48\textwidth}{@{\extracolsep{\fill}}ccc@{}}
\hline\hline
  $|n^{2s_d+1}l_{j_d},NL\rangle$ & $(bc)u$ & $(bc)s$  \\
\hline
 $|1^1s_0, 1S\rangle$  &6.96    &7.11   \\
 $|2^1s_0, 1S\rangle$  &7.32    &7.47   \\
 $|3^1s_0, 1S\rangle$  &7.56    &7.72   \\
 $|4^1s_0, 1S\rangle$  &7.77    &7.92   \\
 $|5^1s_0, 1S\rangle$  &7.96    &8.11   \\
\hline
 $|1^3s_1, 1S\rangle$  &7.00    &7.15   \\
 $|2^3s_1, 1S\rangle$  &7.34    &7.49   \\
 $|3^3s_1, 1S\rangle$  &7.58    &7.73   \\
 $|4^3s_1, 1S\rangle$  &7.79    &7.94   \\
 $|5^3s_1, 1S\rangle$  &7.97    &8.12   \\
\hline
 $|1^1s_0, 1S\rangle$  &6.97    &7.12   \\
 $|1^1p_1, 1S\rangle$  &7.24    &7.39   \\
 $|1^1d_2, 1S\rangle$  &7.42    &7.58   \\
 $|1^1f_3, 1S\rangle$  &7.59    &7.74   \\
 $|1^1g_4, 1S\rangle$  &7.73    &7.89   \\
 $|1^1h_5, 1S\rangle$  &7.87    &8.02   \\
\hline
 $|1^3s_1, 1S\rangle$  &7.00    &7.15   \\
 $|1^3p_2, 1S\rangle$  &7.25    &7.41   \\
 $|1^3d_3, 1S\rangle$  &7.44    &7.59   \\
 $|1^3f_4, 1S\rangle$  &7.60    &7.75   \\
 $|1^3g_5, 1S\rangle$  &7.74    &7.90   \\
 $|1^3h_6, 1S\rangle$  &7.88    &8.03   \\
\hline\hline
\end{tabular*}
\end{table}

\begin{figure*}[!phtb]
\centering
\subfigure[]{\label{subfigure:cfa}\includegraphics[scale=0.48]{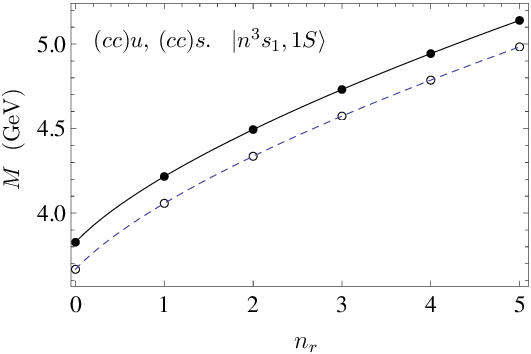}}
\subfigure[]{\label{subfigure:cfa}\includegraphics[scale=0.48]{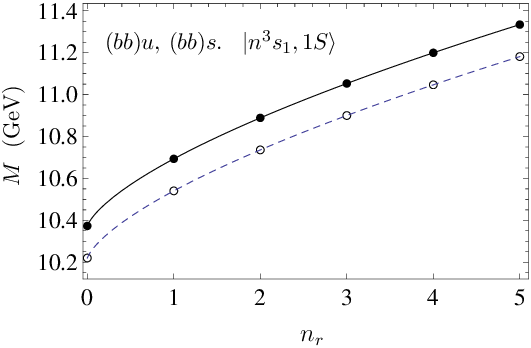}}
\subfigure[]{\label{subfigure:cfa}\includegraphics[scale=0.48]{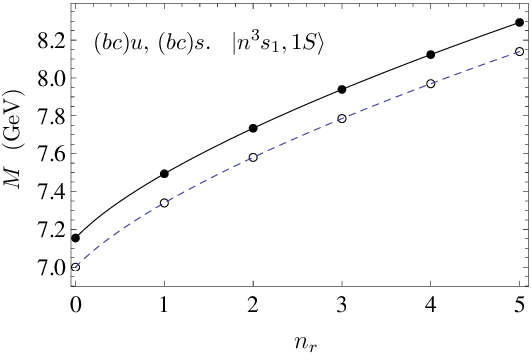}}
\subfigure[]{\label{subfigure:cfa}\includegraphics[scale=0.48]{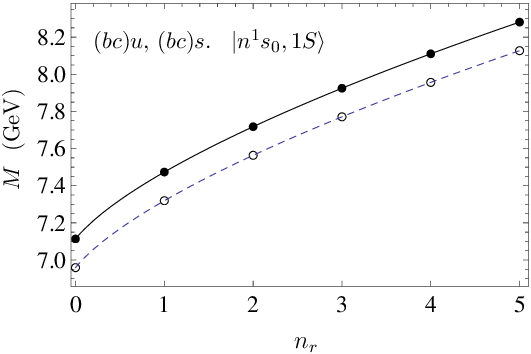}}
\subfigure[]{\label{subfigure:cfa}\includegraphics[scale=0.48]{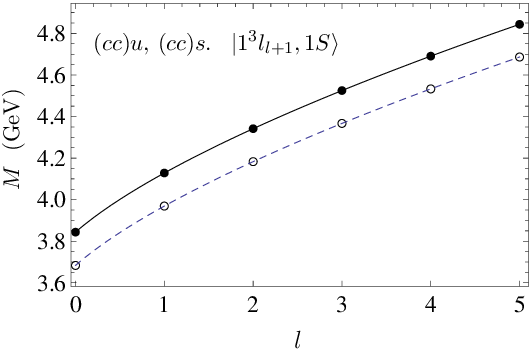}}
\subfigure[]{\label{subfigure:cfa}\includegraphics[scale=0.48]{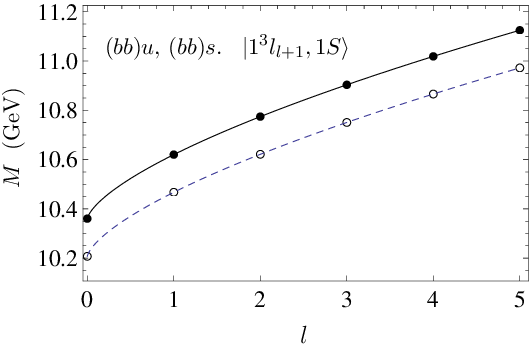}}
\subfigure[]{\label{subfigure:cfa}\includegraphics[scale=0.48]{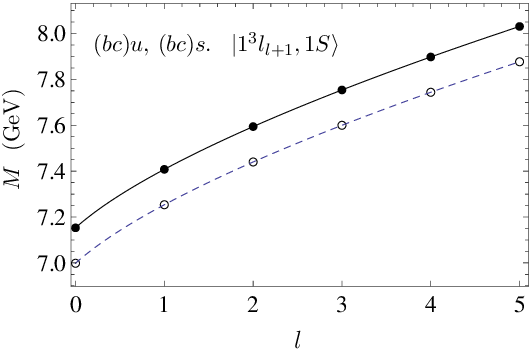}}
\subfigure[]{\label{subfigure:cfa}\includegraphics[scale=0.48]{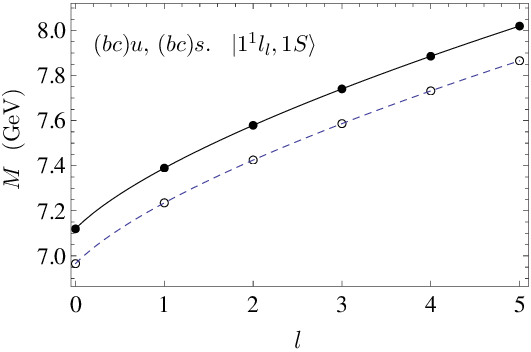}}
\caption{The radial and orbital $\rho$-trajectories for the {\dhbs} $(QQ')q$. Circles and the dashed lines are for the baryons $(QQ')u$ containing an up quark. Dotes and the black lines are for the baryons $(QQ')s$ containing a strange quark. $M$ is the baryon mass. $n=n_r+1$, $n_r$ is the radial quantum number for the $\rho$-mode. $l$ is the orbital quantum number for the $\rho$-mode. Data are listed in Tables \ref{tab:massrho} and \ref{tab:massrhobc}.}\label{fig:rho}
\end{figure*}

\begin{table*}[!phtb]
\caption{Comparison of theoretical predictions for the spin averaged masses of the {\dhbs} (in GeV).}  \label{tab:masscomp}
\centering
\begin{tabular*}{1.0\textwidth}{@{\extracolsep{\fill}}ccccccccc@{}}
\hline\hline
  Baryon  &  $|n^{2s_d+1}l_{j_d},NL\rangle$
          &   Our  & \cite{Ebert:2002ig} &  \cite{Gershtein:2000nx}
          &  \cite{Eakins:2012jk} & \cite{Lu:2017meb} & \cite{Mohanta:2019mxo} &\cite{Shah:2017liu,Shah:2016vmd} \\
\hline
$\Xi_{cc}$ &$|1^3s_1, 1S\rangle$ &3.67 &3.69 &3.57 &3.73 &3.65 & &3.63\\
           &$|1^3s_1, 1P\rangle$ &4.14 &4.14 &4.03 &4.08 &4.02 & &\\
           &$|2^3s_1, 1S\rangle$ &4.06 &3.99 &3.90 &4.06 &4.03 &&3.96\\
$\Xi_{bb}$  &$|1^3s_1, 1S\rangle$&10.22 &10.23&10.12 &10.34 &10.16 &10.09 &10.33\\
           &$|1^3s_1, 2S\rangle$ &10.85 &10.85 &     &10.96 &10.67 &&\\
           &$|1^3s_1, 1P\rangle$ &10.66 &10.67 &10.57&10.69 &10.53 & &\\
           &$|2^3s_1, 1S\rangle$ &10.54 &10.47 &10.40&10.57 &10.47 &&10.61\\
$\Xi_{bc}$ &$|1^1s_0, 1S\rangle$ &6.96 &6.96 &6.85   &      &      &6.79 \\
           &$|1^3s_1, 1S\rangle$ &7.00 &6.96 &       &7.05  &      &6.84 &6.96\\
           &$|1^3s_1, 1P\rangle$ &7.45 &     &       &7.40  &  &\\
           &$|2^3s_1, 1S\rangle$ &7.34 &     &       &7.34  &  &&7.26\\
$\Omega_{cc}$ &$|1^3s_1, 1S\rangle$ &3.83 &3.84   &  &      &3.75 &&3.76\\
           &$|1^3s_1, 1P\rangle$ &4.28 &4.28      &  &      &4.11 & \\
           &$|2^3s_1, 1S\rangle$ &4.22 &4.14      &  &      &4.13 &&4.08\\
$\Omega_{bb}$ &$|1^3s_1, 1S\rangle$ &10.37 &10.38 &  &      &10.25 &10.19&10.46\\
              &$|1^3s_1, 2S\rangle$ &10.98 &10.98 &  &      &10.76 &\\
           &$|1^3s_1, 1P\rangle$ &10.79 &10.80    &  &      &10.61 &\\
           &$|2^3s_1, 1S\rangle$ &10.69 &10.63    &  &      &10.57 &&10.72\\
$\Omega_{bc}$&$|1^1s_0, 1S\rangle$ &7.11 &7.12      &  &      &       &6.89 \\
           &$|1^3s_1, 1S\rangle$ &7.15 &7.12    &  &      &       &6.95&7.17\\
           &$|1^3s_1, 1P\rangle$ &7.58 &          \\
           &$|2^3s_1, 1S\rangle$ &7.49 & &&&&&7.49\\
\hline\hline
\end{tabular*}
\end{table*}

\subsection{$\lambda$- and $\rho$-trajectories for baryons $(QQ')q$}\label{subsec:rtsa}

When calculating the $\lambda$-mode radially and orbitally excited states, the $\rho$-mode state is taken as the radial ground state. Similarly, when calculating the $\rho$-mode radially and orbitally excited states, the $\lambda$-mode state is taken as the radial ground state.

Using Eqs. (\ref{t2q}), (\ref{pa2qQ}), and (\ref{fitcfxl}) [or Eqs. (\ref{combrt}), (\ref{pa2qQ}), and (\ref{fitcfxl})], along with parameters in Table \ref{tab:parmv}, the spectra of the $\lambda$-excited states and the $\rho$-excited states of the {\dhbs} $(QQ')q$ can be calculated, see Tables \ref{tab:masslambda}, \ref{tab:masslambdabc}, \ref{tab:massrho} and \ref{tab:massrhobc}.
The radial and orbital $\lambda$- and $\rho$-trajectories for the {\dhbs} $(cc)q$, $(bb)q$ and $(bc)q$ are illustrated in Figs. \ref{fig:brho} and \ref{fig:rho}.

We assume $m_u=m_d$; therefore, the masses of $(QQ')u$ and $(QQ')d$ are equal, and only the masses of $(QQ')u$ are presented. When calculating baryon masses, the masses of diquarks $(QQ')$ are computed using the diquark {\rt} relation [the second line in Eq. (\ref{t2q}) and Eq. (\ref{pa2qQ})]. For the diquarks $(cc)$ and $(bb)$, the states $1^3p_2$, $1^3f_4$, and $1^3h_6$ do not exist due to the Pauli exclusion principle \cite{Feng:2023txx}. Consequently, the corresponding states of the {\dhbs} also do not exist, see Table \ref{tab:massrho}.

As discussed in Appendix \ref{sec:appcfx}, the values of $c_{fx_{\lambda}}$ and $c_{0x_{\lambda}}$ vary with the masses of diquark and quark, $M_{\rho}$ and $m_q$, as shown in Eqs. (\ref{pa2qQ}) and (\ref{fitcfxl}). This indicates that $c_{fx_{\lambda}}$ and $c_{0x_{\lambda}}$ change with different diquark states when discussing the $\rho$-trajectories, even though the $\lambda$-mode remains unchanged.

The obtained results for both the $\lambda$-excited states and $\rho$-excited states are consistent with other theoretical predictions, see Table \ref{tab:masscomp}.
The mass of the observed {\dhb} $\Xi_{cc}^{+}$ \cite{SELEX:2002wqn,SELEX:2004lln} is $3518.9\pm0.9$ {\mev} \cite{ParticleDataGroup:2024cfk}. The mass of $\Xi^{++}_{cc}$ \cite{LHCb:2017iph,LHCb:2018pcs,LHCb:2019epo} is $3621.6\pm0.4$ {\mev} \cite{ParticleDataGroup:2024cfk}. It is found that the experimental data agree with the theoretical predictions for the ground state of $\Xi_{cc}$, see Table \ref{tab:masscomp}; therefore, it is expected that the observed states $\Xi_{cc}^{+}(3519)$ and $\Xi^{++}_{cc}(3622)$ are the ground states of the {\dhb} $\Xi_{cc}$.

\subsection{Discussions}

By employing the baryon {\rt} relations in Eqs. (\ref{t2q}) and (\ref{pa2qQ}) or Eqs. (\ref{combrt}) and (\ref{pa2qQ}), the masses of highly excited states can be calculated easily. While highly excited states are expected to be unstable, the calculation of their masses is still instructive. This is because various models or approaches use different parameter values to calculate the masses of the {\dhbs}. Despite using different parameter values, these models often provide consistent predictions for the ground state and lower excited states. The theoretical predictions for the highly excited states are expected to show differences, which will be instructive.

The Regge trajectories take different form and behave differently in various energy regions \cite{Chen:2022flh,Chen:2021kfw}. For the $\lambda$-mode, the {\dhbs} $(QQ')q$ is a heavy-light system; therefore, the $\lambda$-trajectories behave as $M{\sim}x^{1/2}_{\lambda}$, see Eq. (\ref{combrt}). In contrast, for the $\rho$-mode, the diquark $(QQ')$ in the {\dhbs} $(QQ')q$ is a heavy-heavy system, causing the $\rho$-trajectories to behave as $M{\sim}x^{2/3}_{\rho}$.

Corresponding to two series {\rts}, there are two series of masses: one for the $\lambda$-excited states and the other for the $\rho$-excited states. There is no degeneracy in masses, see Tables \ref{tab:masslambda}, \ref{tab:masslambdabc}, \ref{tab:massrho} and \ref{tab:massrhobc}. For an excited state, the mass of the $\lambda$-mode will be larger than that of the $\rho$-mode. In this work, different mixings are not considered, such as the $^3(j-1)_j-^3(j+1)_j$ mixing of the doubly heavy diquarks, which is analogous to the mixing of different wave states of charmonium and bottomonium \cite{Eichten:2007qx}.

\section{Conclusions}\label{sec:conc}
In this work, by applying the diquark {\rt} relation, we present the explicit form of the Regge trajectory relations for the doubly heavy baryons $\Xi_{cc}$, $\Xi_{bc}$, $\Xi_{bb}$, $\Omega_{cc}$, $\Omega_{bc}$ and $\Omega_{bb}$ in the diquark picture.
By employing the obtained Regge trajectory relations, the masses of the $\lambda$-excited states and the $\rho$-excited states are estimated and shown to be consistent with other theoretical predictions.

Both the $\lambda$-trajectories and the $\rho$-trajectories are presented. For the {\dhbs}, it is demonstrated that the $\rho$-trajectories behave differently from the $\lambda$-trajectories. Specifically, the $\rho$-trajectories behave as $M{\sim}x_{\rho}^{2/3}$ $(x_{\rho}=n_r,l)$, whereas the $\lambda$-trajectories behave as $M{\sim}x_{\lambda}^{1/2}$ $(x_{\lambda}=N_r,L)$.
By utilizing the obtained relations, the baryon Regge trajectory provides a very simple approach for estimating the spectra of both the $\lambda$-excited states and the $\rho$-excited states.

\appendix
\section{States of baryons}
The states of baryons in the diquark picture are listed in Table \ref{tab:bqstates}.

\begin{table*}[!phtb]
\caption{The states of baryons composed of one diquark in $\cltba$ and one quark in $3_c$. The notation is explained in \ref{subsec:prelim}. Here, $q$, $\qp$ and $\qpp$ represent both the light quarks and the heavy quarks.}  \label{tab:bqstates}
\centering
\begin{tabular*}{\textwidth}{@{\extracolsep{\fill}}ccccc@{}}
\hline\hline
$J^P$ & $(L,l)$  &  Configuration \\
\hline
$\frac{1}{2}^+$ & $(0,0)$ & $\left({\qqs}^{\cltba}_{n^1s_0}{\qpp}\right)_{N^2S_{1/2}}$,\;
$\left({\qqb}^{\cltba}_{n^3s_1}{\qpp}\right)_{N^2S_{1/2}}$,\\
& $(1,1)$ &
$\left({\qqb}^{\cltba}_{n^1p_1}{\qpp}\right)_{N^2P_{1/2}}$,\;
$\left({\qqb}^{\cltba}_{n^1p_1}{\qpp}\right)_{N^4P_{1/2}}$,\;
$\left({\qqs}^{\cltba}_{n^3p_0}{\qpp}\right)_{N^2P_{1/2}}$,\;
$\left({\qqs}^{\cltba}_{n^3p_1}{\qpp}\right)_{N^2P_{1/2}}$,\\
& &
$\left({\qqs}^{\cltba}_{n^3p_1}{\qpp}\right)_{N^4P_{1/2}}$,\;
$\left({\qqs}^{\cltba}_{n^3p_2}{\qpp}\right)_{N^4P_{1/2}}$\\
&$\cdots$ &$\cdots$ \\
$\frac{1}{2}^-$ & $(1,0)$ &
$\left({\qqs}^{\cltba}_{n^1s_0}{\qpp}\right)_{N^2P_{1/2}}$,\;
$\left({\qqb}^{\cltba}_{n^3s_1}{\qpp}\right)_{N^2P_{1/2}}$,\;
$\left({\qqb}^{\cltba}_{n^3s_1}{\qpp}\right)_{N^4P_{1/2}}$,\\
& $(0,1)$ &
$\left({\qqb}^{\cltba}_{n^1p_1}{\qpp}\right)_{N^2S_{1/2}}$,\;
$\left({\qqs}^{\cltba}_{n^3p_0}{\qpp}\right)_{N^2S_{1/2}}$,\;
$\left({\qqs}^{\cltba}_{n^3p_1}{\qpp}\right)_{N^2S_{1/2}}$,\\
&$\cdots$ &$\cdots$ \\
$\frac{3}{2}^+$ & $(0,0)$ &
$\left({\qqb}^{\cltba}_{n^3s_1}{\qpp}\right)_{N^4S_{3/2}}$,\\
& $(1,1)$ &
$\left({\qqb}^{\cltba}_{n^1p_1}{\qpp}\right)_{N^2P_{3/2}}$,\;
$\left({\qqb}^{\cltba}_{n^1p_1}{\qpp}\right)_{N^4P_{3/2}}$,\;
$\left({\qqs}^{\cltba}_{n^3p_0}{\qpp}\right)_{N^2P_{3/2}}$,\;
$\left({\qqs}^{\cltba}_{n^3p_1}{\qpp}\right)_{N^2P_{3/2}}$,\;\\
& &
$\left({\qqs}^{\cltba}_{n^3p_1}{\qpp}\right)_{N^4P_{3/2}}$,\;
$\left({\qqs}^{\cltba}_{n^3p_2}{\qpp}\right)_{N^4P_{3/2}}$,\;
$\left({\qqs}^{\cltba}_{n^3p_2}{\qpp}\right)_{N^6P_{3/2}}$,\\
&$\cdots$ &$\cdots$ \\
$\frac{3}{2}^-$ & $(1,0)$ &
$\left({\qqs}^{\cltba}_{n^1s_0}{\qpp}\right)_{N^2P_{3/2}}$,\;
$\left({\qqb}^{\cltba}_{n^3s_1}{\qpp}\right)_{N^2P_{3/2}}$,\;
$\left({\qqb}^{\cltba}_{n^3s_1}{\qpp}\right)_{N^4P_{3/2}}$,\\
& $(0,1)$ &
$\left({\qqb}^{\cltba}_{n^1p_1}{\qpp}\right)_{N^4S_{3/2}}$,\;
$\left({\qqs}^{\cltba}_{n^3p_1}{\qpp}\right)_{N^4S_{3/2}}$,\;
$\left({\qqs}^{\cltba}_{n^3p_2}{\qpp}\right)_{N^4S_{3/2}}$\\
&$\cdots$ &$\cdots$ \\
\hline\hline
\end{tabular*}
\end{table*}

\section{Determination of $c_{fx_{\lambda}}$ and $c_{0x_{\lambda}}$ for the $\lambda$-modes of the heavy-light systems}\label{sec:appcfx}

In this section, we determine the values of $c_{fx_{\lambda}}$ and $c_{0x_{\lambda}}$ for the $\lambda$-modes of the {\dhbs} $(QQ')q$.
Eq. (\ref{massfinal}) is applied to fit the {\rts} for the heavy-light mesons and the $\lambda$-modes of heavy-light baryons, which are composed of one doubly heavy diquark and one light quark.
The quality of a fit is measured by the quantity $\chi^2$, defined as
\bea
\chi^2=\frac{1}{N-1}\sum^{N}_{i=1}\left(\frac{M_{fi}-M_{ei}}{M_{ei}}\right)^2,
\eea
where $N$ is the number of points on the trajectory, $M_{fi}$ is fitted value, and $M_{ei}$ is the experimental value or the theoretical value of the $i$-th particle mass. The parameters are determined by minimizing $\chi^2$.

\begin{table}[!phtb]
\caption{The spin averaged masses of the radially excited states of mesons and baryons (in ${\gev}$).}  \label{tab:cfr}
\centering
\begin{tabular*}{0.5\textwidth}{@{\extracolsep{\fill}}ccccccc@{}}
\hline\hline
           & $1S$  &  $2S$  &  $3S$  & $4S$  &  $5S$  & $6S$  \\
\hline
$c\bar{u}$  &1.97135 &2.6075  &3.0875  &3.4745  &3.81475   \\
$c\bar{s}$  &2.07624 &2.68325 &3.23625 &3.66475 &4.04425  \\
$b\bar{u}$  &5.31342 &5.902  &6.385  &6.78475 &7.132  \\
$b\bar{s}$  &5.40328 &5.988  &6.473  &6.87775  &7.234   \\
$\Xi_{bb}$  &10.2253   &10.8507  &  &  &   \\
$\Omega_{bb}$  &10.379   &10.9847  &  &  &   \\
$\Lambda_{c}$  &2.286   &2.769  &3.130  &3.437  &3.715 &3.973  \\
$\Lambda_{b}$  &5.620   &6.089  &6.455  &6.756  &7.015 &7.256  \\
$\Xi_{c}$  &2.476   &2.959  &3.323  &3.632  &3.909 &4.166   \\
$\Xi_{b}$  &5.803   &6.266  &6.601  &6.913  &7.165 &7.415   \\
\hline\hline
\end{tabular*}
\end{table}

\begin{table}[!phtb]
\caption{The spin averaged masses of the orbitally excited states of mesons and baryons (in ${\gev}$).}  \label{tab:cfo}
\centering
\begin{tabular*}{0.5\textwidth}{@{\extracolsep{\fill}}ccccccc@{}}
\hline\hline
            & $1S$  &  $1P$  &  $1D$  & $1F$  &  $1G$ &$1H$   \\
\hline
$c\bar{u}$  &1.97135   &2.42923  &2.83425  &3.14468  &3.41747   \\
$c\bar{s}$  &2.07624   &2.51226  &2.9498  &3.2675  &3.55286   \\
$b\bar{u}$  &5.31342   &5.74542  &6.1057  &6.39846  &6.64783   \\
$b\bar{s}$  &5.40328   &5.84425  &6.19995  &6.48789  &6.7375   \\
$\Xi_{cc}$  &3.69133   &4.13867  &  &  &   \\
$\Xi_{bb}$  &10.2253   &10.6636  &  &  &   \\
$\Omega_{cc}$  &3.84067   &4.28244  &  &  &   \\
$\Omega_{bb}$  &10.379   &10.7979  &  &  &   \\
$\Lambda_{c}$  &2.286   &2.61733  &2.8776  &3.08614  &3.27778 &3.45273  \\
$\Lambda_{b}$  &5.620   &5.938  &6.1936  &6.40971  &6.59856 &6.76645   \\
$\Xi_{c}$  &2.476   &2.81  &3.0692  &3.286  &3.47678 &3.65118   \\
$\Xi_{b}$  &5.803   &6.12667  &6.3702  &6.57929  &6.76111 &6.93355   \\
\hline\hline
\end{tabular*}
\end{table}

The masses used are listed in Tables \ref{tab:cfr} and \ref{tab:cfo}. For experimentally determined states, the PDG data from Ref. \cite{ParticleDataGroup:2024cfk} are used. For baryons and the undetermined states of mesons, theoretical data from Refs. \cite{Ebert:2009ua,Ebert:2002ig,Ebert:2011kk} are used.
In addition to the masses of bound states, some parameters are provided in Table \ref{tab:parmv}. The masses of scalar diquarks are $m_{[ud]}=0.68$ ${\gev}$, $m_{[us]}=0.90$ ${\gev}$ \cite{Chen:2023ngj}. The masses of diquarks $(QQ')$ ($Q,Q'=b,c$) are calculated using the parameters in Table \ref{tab:parmv} along with Eqs. (\ref{t2q}) and (\ref{pa2qQ}) \cite{Feng:2023txx}.

\begin{figure}[!phtb]
\centering
\subfigure[]{\label{subfigure:cfa}\includegraphics[scale=0.7]{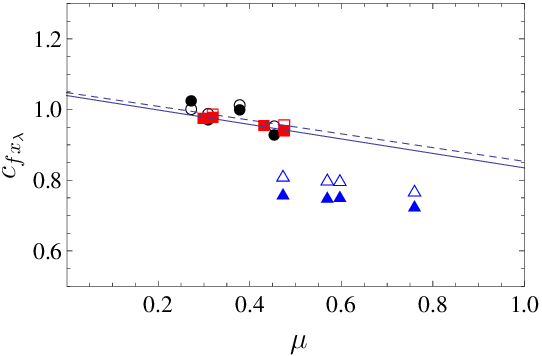}}
\subfigure[]{\label{subfigure:cfb}\includegraphics[scale=0.7]{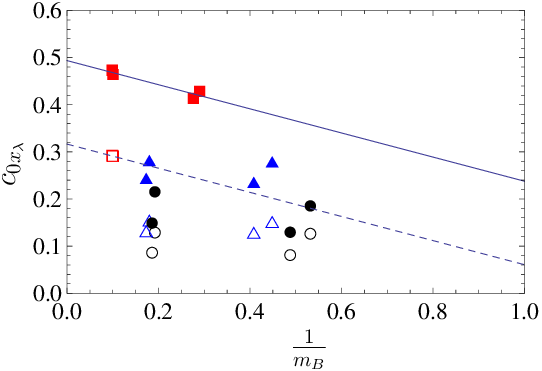}}
\caption{The fitted $c_{fx_{\lambda}}$ [Fig. \ref{subfigure:cfa}] and $c_{0x_{\lambda}}$ [Fig. \ref{subfigure:cfb}], ($x_{\lambda}=L,\,N_r$). The data are listed in Table \ref{tab:fp}. $\mu$ and $m_B$ are listed in Eq. (\ref{dhbrm}). The black filled circles (mesons), red filled squares (doubly heavy baryons), and blue filled triangles (singly heavy baryons) correspond to the fitted $c_{fL}$ or $c_{0L}$. Similarly, the black empty circles (mesons), red empty squares (doubly heavy baryons), and blue empty triangles (singly heavy baryons) correspond to the fitted $c_{fN_r}$ or $c_{0N_r}$. The black lines ($c_{fL}$ and $c_{0L}$) and the dashed lines ($c_{fN_r}$ and $c_{0N_r}$) represent the linear fit for the doubly heavy baryons, see Eq. (\ref{fitcfxl}). }\label{fig:rgc0fx}
\end{figure}

Using Eq. (\ref{massfinal}) with $\sigma'=C$, the fitted values are obtained, see Table \ref{tab:fp} and Fig. \ref{fig:rgc0fx}.
In Fig. \ref{fig:rgc0fx}, $c_{fx_{\lambda}}$ and $c_{0x_{\lambda}}$ for the heavy-light mesons and the singly heavy baryons are also depicted to show explicitly the similar relations.
According to Eq. (\ref{t2q}), $c_{fx_{\lambda}}$ and $c_{0x_{\lambda}}$ are required to determine a {\rt}, as $m_{R_{\lambda}}$ can be calculated by using Eq. (\ref{pa2qQ}) and parameters in Table \ref{tab:parmv}.
Two or more states on the {\rt} are needed to obtain $c_{fx_{\lambda}}$ and $c_{0x_{\lambda}}$. These parameters are determined by fitting the data from the doubly heavy baryons. From the data in Table \ref{tab:fp}, the fitted relations are:
\begin{eqnarray}\label{fitcfxl}
c_{fL}=&1.040 - 0.205\mu,\; c_{0L}=0.494 - \frac{0.256}{m_B}, \nonumber\\
c_{fN_r}=&1.048 - 0.195\mu, \;  c_{0N_r}=0.316-\frac{0.256}{m_B},
\end{eqnarray}
where
\bea\label{dhbrm}
m_B=M_{{\rho}}+m_q,\quad \mu=M_{\rho}m_q/(M_{{\rho}}+m_q).
\eea
Here, $\mu$ is the reduced masses for the doubly heavy baryons, see Eq. (\ref{dhbrm} ). $M_{{\rho}}$ is mass of the doubly heavy diquark, see Eq. (\ref{pa2qQ}). As additional experimental or theoretical data become available, the fitted formulas will be further refined.

When fitting $c_{fL}$ and $c_{0L}$, $\Xi_{cc}$, $\Xi_{bb}$, $\Omega_{cc}$ and $\Omega_{bb}$ are considered. When fitting $c_{fN_r}$ and $c_{0N_r}$, only the doubly bottom baryons $\Xi_{bb}$ and $\Omega_{bb}$ are considered. When fitting $c_{0N_r}$, we assume that the slopes of $c_{0L}$ and $c_{0N_r}$ in Fig. \ref{subfigure:cfb} are equal. The averaged value of $c_{0N_r}$ for $\Xi_{bb}$ and $\Omega_{bb}$ is then used to obtain the intercept $ 0.316$ in Eq. (\ref{fitcfxl}).

\begin{table}[!phtb]
\caption{The fitted $c_{fx_{\lambda}}$ and $c_{0x_{\lambda}}$.}  \label{tab:fp}
\centering
\begin{tabular*}{0.5\textwidth}{@{\extracolsep{\fill}}ccc@{}}
\hline\hline
          & $(c_{fN_r},\,c_{0N_r})$  & $(c_{fL},\,c_{0L})$ \\

$c\bar{u}$  & $(1.00097, 0.126205)$  & $(1.02452, 0.185294)$  \\
$c\bar{s}$  & $(1.01242, 0.0809647)$  & $(0.999118, 0.129391)$ \\
$b\bar{u}$  & $(0.987851, 0.128748)$  & $(0.970877, 0.215201)$ \\
$b\bar{s}$  & $(0.952508, 0.0859585)$  & $(0.927975, 0.148784)$ \\
$\Xi_{cc}$  &                          & $(0.9749, 0.428312)$ \\
$\Xi_{bb}$  & $(0.985851, 0.290274)$  & $(0.977695, 0.463893)$ \\
$\Omega_{cc}$&                         &$(0.954013, 0.413617)$  \\
$\Omega_{bb}$&$(0.955422, 0.291725)$   &$(0.93996, 0.47324)$  \\
$\Lambda_{c}$&$(0.812679, 0.151027)$   &$(0.761658, 0.27892)$  \\
$\Lambda_{b}$&$(0.800169, 0.15407)$   &$(0.754373, 0.281716)$  \\
$\Xi_{c}$    &$(0.801898, 0.128107)$   &$(0.752152, 0.235414)$  \\
$\Xi_{b}$    &$(0.770306, 0.131342)$   &$(0.727442, 0.243779)$  \\
\hline\hline
\end{tabular*}
\end{table}

\end{document}